# High-yield production of 2D crystals by wet-jet milling

A. E. Del Rio Castillo[a]*, V. Pellegrini[a,b], A. Ansaldo[a], F. Ricciardella[a], H. Sun[a], L. Marasco[a], J. Buha[c], Z. Dang[c], L. Gagliani[a], E. Lago[a], N. Curreli[a], S. Gentiluomo[a], F. Palazon[c], P. Toth[a], E. Mantero[a], M. Crugliano[a], A. Gamucci[a,b], A. Tomadin[a], M. Polini[a], and F. Bonaccorso[a,b]*

[a] Istituto Italiano di Tecnologia, Graphene Labs, Via Morego 30, 16163 Genova, Italy.

[b] BeDimensional Srl., Via Albisola 121, 16163 Genova, Italy.

[c] Istituto Italiano di Tecnologia, Nanochemistry Department, Via Morego 30, 16163 Genova, Italy.

* Corresponding authors: antonio.delrio@iit.it, francesco.bonaccorso@iit.it.

## Abstract

Efficient and scalable production of two-dimensional (2D) materials is required to overcome technological hurdles towards the creation of a 2D-materials-based industry. Here, we present a novel approach developed for the exfoliation of layered crystals, *i.e.*, graphite, hexagonal-boron nitride and transition metal dichalcogenides. The process is based on high-pressure wet-jet-milling (WJM), resulting in 2 L hr$^{-1}$ production of 10 gL$^{-1}$ of single- and few-layer 2D crystal flakes in dispersion making the scaling-up more affordable. The WJM process enables the production of defect-free and high quality 2D-crystal dispersions on a large scale, opening the way for the full exploitation in different commercial applications, *e.g.*, anodes active material in lithium ion batteries, reinforcement in polymer-graphene composites, and transparent conductors as we demonstrate in this report.

## Introduction

Since the isolation and characterization of graphene in 2005,[1] its possible applications are increasing year by year.[2,3,4,5,6,7,8] Graphene promises to revolutionize the plastic market, providing extra-properties to polymer composites, *i.e.*, increasing their mechanical properties[9,10,11,12] and enhancing the electrical[9,10,11] and thermal conductivities.[13,14,15,16] Additionally, its applications in the energy and (opto)electronics fields are extensive, covering a wide range of energy storage[17,18,19,20,21] and energy production devices,[22,23] sensors,[24,25,26,27,28,29,30] high speed transistors,[31,32,33] photodetectors,[34,35,36,37] modulators[38,39,40] and mode locking lasers.[41,42,43,44] Despite the several applications in which graphene can potentially play a key role, currently a large-scale synthesis process, compatible with the industrial requirements of mass production and repeatability, is still lacking. To this end we recall that the synthesis of graphene relies on two main routes: the bottom-up and the top-down approaches.[45] The chemical vapour deposition (CVD) is the most representative and industrially-relevant bottom-up technique.[8,45,46,47] Graphene grown by CVD is characterized by high quality, large grain size – up to one centimetre,[48] and despite its higher cost it is suitable for high value-added applications, *e.g.*, photonics,[49] electronics[50,51] and flexible electronics.[8,45,47]

On the contrary, in the top-down approach, graphite crystals are exfoliated or peeled-off to achieve ultra-thin flakes.[1,2] The most commonly-used top-down methods are the micromechanical cleavage (MC)[52] and the liquid phase exfoliation (LPE).[53,56] The MC, consisting in the consecutively peeling-off of graphite flakes by using an adhesive tape, provides high quality flakes in terms of crystallinity and morphology, *i.e.*, lateral size and thickness.[54] However, MC is mostly suitable for fundamental studies and the realization of new concept devices,[52] but it is impractical for large-scale production.[45,52,55] In the LPE method, graphite is exfoliated in liquid solvents by exploiting cavitation[56,57,58,59,60,61,62,] or shear forces[63,64,65,66] to extract single- (SLG) and few-layers (FLG) graphene flakes. The LPE process can be scaled up and the exfoliated flakes can be deposited or printed on different substrates using well-known techniques *e.g.*, ink-jet printing, flexography, spray-coating.[45] Generally, LPE consists of three mains steps: dispersion, exfoliation, and purification,[45] where the exfoliation step is commonly performed by ultra-sonication[58,59,60,61,62,65] or high-shear mixing,[66,67,68,69,70] while the purification is carried out by means of ultracentrifugation.[45,56,71]

In order to evaluate the effectiveness of LPE techniques and compare the effectiveness of the exfoliation in terms of production rate and time required for the exfoliation, it is necessary to establish a set of figures of merit (FoM). For example, numbers of SLG versus the total number of flakes present in the sample, or the SLG mass fraction were proposed as FoM.[56] However, since their determination is tedious and time consuming, they have been seldom used.[45] Instead a FoM which is largely used is the exfoliation yield by weight -$Y_W$ [%]-, *i.e.*, the ratio between the weight of the final graphitic material and the weight of the starting graphite flakes.[45] Additionally to the above introduced FoM, and in view of the large-scale production of high quality 2D crystals (*i.e.*, single- and few-layer flakes), here we propose to set the 1 g of exfoliated 2D crystals as a standard for the definition of two further FoM. The first one is the time required to obtain 1 g of exfoliated 2D crystals in dispersion after the exfoliation process, $t_{1gram}$ [min], and the second one is the volume of solvent required to produce 1 g of exfoliated 2D crystals, $V_{1gram}$ [L]. The quantity $V_{1gram}$ is calculated directly from the concentration of exfoliated flakes in suspension and $Y_W$. $V_{1gram}$ is thus a direct tool to evaluate the amount of solvent required for the production of 2D crystals, which is an important factor in view of large-scale production.

To evaluate the effectiveness of the LPE process, we focus on the techniques that are most commonly used and promising in terms of scalability, *i.e.*, (1) ultra-sonication, (2) ball-milling, (3) shear-exfoliation, and (4) micro-fluidization.

(1) Exfoliation by ultra-sonication is the most widely-exploited LPE technique due to the ease and simplicity of the process.[72] The creation of cavitation during ultra-sonication[73] induces the exfoliation of layered crystals.[57] For a typical ultra-sonication process of 6 hours the FoM values are, $t_{1gram}$ > 360 min, $V_{1gram} \approx$ 3.3 L and $Y_W \approx$ 3%.[56-62,73] To the best of our knowledge, the highest value of $C$ obtained by ultra-sonication is 60 gL$^{-1}$,[74] with $Y_W \approx$ 19%, obtained after more than 35 hours of sonication ($t_{1gram}$ > 1800) and several steps of precipitation by ultracentrifugation and re-dispersion ($V_{1gram} \approx$ 0.53 L). (2) The planetary ball-milling method consists in mixing graphite and solvent in a planetary-rotatory mill.[63,64,75] The zirconia or metallic container (jar) is filled with balls of the same material. During the jar spinning, the crashing and friction between balls creates shear forces promoting the exfoliation of graphite.[63,64,75] Exploiting the exfoliation of graphite by planetary ball-milling, $C$ close to 0.2 gL$^{-1}$,[63,64] for processing times ranging from 1 to 30 hours, have been obtained.[63,64] The values of the FoM are $Y_W$ < 1%,[63] $t_{1gram} \approx$ 60 min and $V_{1gram} \approx$ 100 L.[63,64] However, the scalability and repeatability of this method have not been proved yet. (3) The shear mixer has emerged as a new tool for the exfoliation of layered crystals.[66,67] Turbulences and shear forces produced by the rotor/stator reciprocal motion exfoliate the layered crystals in dispersion.[66] The value of $C$ reached by means of high-shear mixing is demonstrated to be 0.1 gL$^{-1}$ after more than 60 hours of process.[66] Albeit the large volumes (hundreds of L) processed by shear exfoliation, the $Y_M \approx$ 0.002%, $t_{1gram} \approx$ 3600 min and $V_{1gram} \approx$ 10 L,[66] make the shear exfoliation a technique still to be improved to fulfil the industrial scale demand. (4) A promising technique recently reported is micro-fluidization,[76,77,78] which consists in subjecting the layered crystals dispersion to high shear rates (10$^8$ s$^{-1}$).[78] By exploiting this technique, a $C$ of 100 gL$^{-1}$ and $Y_M \approx$ 100% have been reached.[78] The processing time is limited to the piston passes (70 passes), meaning $t_{1gram} \approx$ 115-230 min and $V_{1gram} \approx$ 0.18-0.36 L.[78] This technique has been demonstrated using water/surfactant only and the exfoliated flakes manifest structural defects, which increase proportionally with the piston passes.[78]

Graphite apart, the aforementioned LPE techniques have been applied for the exfoliation of other layered crystals. In particular, micro-fluidization has been used for hexagonal boron nitride (h-BN) flakes,[79] high-shear mixing to exfoliate bulk black phosphorous,[80] MoS$_2$,[81] h-BN,[66] WS$_2$,[66] MoSe$_2$,[66] and MoTe$_2$,[66] ball milling for h-BN,[82,83] and MoS$_2$,[82] while a large number of layered crystals have been exfoliated using ultrasonic bath.[56,84,85,86,108,109]

The exfoliated flakes resulting from the application of these techniques have been used in a number of applications, ranging from polymer composite reinforcement [12,87] to functional inks.[88,89,90] However, to bridge those applications from the lab scale to the market,[8] the development of an affordable production strategy, that allows the production of high quality 2D crystals on large scale, is still required.

In this Article, we propose the use of the wet-jet mill (WJM) process to exfoliate different layered crystals, *i.e.*, graphite, *h*-BN, $MoS_2$ and $WS_2$ for the large-scale production of high quality 2D crystals. The WJM exploits high pressure (180-250 MPa) to force the passage of the solvent/layered-crystal mixture through perforated disks, with adjustable hole diameters (0.3-0.1 mm, named nozzle), strongly enhancing the effectiveness of shear forces.[91] The main advantage of WJM compared to all the aforementioned techniques, is the process time of the sample, *i.e.*, the passage of the processed dispersion through the nozzle, which is reduced to a fraction of a second, instead of hours in sonic bath[56-62,73,74] or shear exfoliation.[66,67]

By using WJM, we report here the production of 20 L of 2D crystals dispersed in an organic solvent, with a $C$ of 10 g$L^{-1}$ in 8.5 hours. Considering the case of graphite, we were able to achieve $t_{1gram} \approx 2.55$ min and $V_{1gram} \approx 0.1$ L, with $Y_W$ of 100%. The resulting exfoliated graphene flakes have average thickness of ~3.2 nm and main lateral size distribution of ~500 nm. The exfoliation of *h*-BN, $MoS_2$ and $WS_2$ produces flakes with similar lateral sizes (~500 nm) and thickness (~3 nm). In addition, we demonstrate that 2D crystals obtained by the WJM, *i.e.*, SLG/FLG, are suitable for a range of applications, where large volume of material is needed for the industrial implementation.

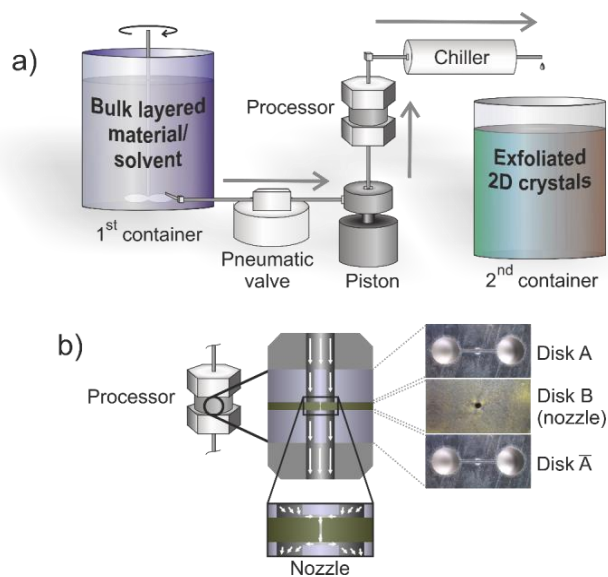

**Figure 1.** (a) Scheme of the wet-jet mill system, the arrows indicate the flow of the solvent through the WJM, (b) Close-up view of the processor. The zoomed image in (b) shows the channels configuration and the disks arrangement. The solvent flow is indicated by the white arrows. On the right side a top view of the holes and channels on each disk. The disks A and Ā have two holes of 1 mm in diameter, separated by a distance of 2.3 mm from centre to centre and joined by a half-cylinder channel of 0.3 mm in diameter. The thickness of the A and Ā disks is 4 mm. The disk B is the core of the system; the image (b, disk B) shows the 0.10 mm nozzle. It can be changed to 0.10, 0.20,

and 0.30 mm nozzle diameter disks according with the size of the bulk layered crystals. The thickness of the B disk is 0.95 mm.

## The wet-jet milling

The WJM apparatus is schematised in Figure 1a. A hydraulic mechanism and a piston supply the pressure (up to 250 MPa) in order to push the sample into a set of 5 different perforated and interconnected disks, see Figure 1b, named processor, where jet streams are generated. The common industrial use of the WJM consists of the pulverization of drugs or paints.[92,93,94] The pulverization is obtained mainly by colliding the pressurised streams of the particle liquid dispersions. The collision takes place between the disks A and B (Figure 1b).[93] In contrast, for the exfoliation of layered crystals, the shear force generated by the solvent when the sample passes through the disk B, as discussed in the WJM modelling section, is the main phenomenon promoting the exfoliation. An important factor that must be considered for the exfoliation of layered crystals is the solvent selection. In fact, in order to exfoliate layered crystals, the Gibbs free energy of the mixture solvent/layered-material must be minimized.[56,95,96,] This condition can be attained if the surface tension ($\gamma$) of the solvent is equivalent to the surface free energy of the material,[56] *i.e.*, as the surface energies of graphite, *h*-BN, $MoS_2$ and $WS_2$ are ~62,[56,88] ~65,[72] ~70,[72] and ~75[72] mJ m$^{-2}$ respectively. N-methyl-2-pyrrolidone, 1.2-dichlorobenzene, or a mixture of ethanol and water[97,98,99] [5:5] having a $\gamma$ = 40.8,[56] 37.0[100] and 30.9[88] mNm$^{-1}$, respectively, can be used as solvents to exfoliate the aforementioned layered crystals. For details on the $\gamma$ and surface energy see the Appendix.

## Wet-jet mill modelling

Among the several liquid-phase approaches to exfoliation, micro-fluidization is the most similar to WJM, in that the whole fluid is forced through a spatial region where the flow becomes turbulent.[76-78] In the micro-fluidization case, such region is a microchannel, while in the WJM it corresponds to the channel junctions before and after the nozzle. In this region, turbulent flow results in a high-shear rate, i.e. velocity gradient orthogonal to the flow direction.

The resulting shear stress applied to the dispersed flakes induces sliding of the 2D crystal planes and initiates the exfoliation process. For graphene, it has been shown that shear rates in excess of $10^4$ s$^{-1}$ are sufficient for the exfoliation process to occur.[66] These values can be achieved in the laminar flow produced by shear-mixers,[66] and, more efficiently, in the turbulent flow of micro-fluidizers[78] and WJMs.

A most salient difference between the exfoliation process in the WJM and other LPE methods aforementioned is the large pressure drop in time experienced by the crystallites as the dispersion flows through the nozzle, specifically through the disk B. In the following we attribute the high production yield of the WJM to a geometry-induced enhancement of the shear exfoliation rate since we must rule out that the large pressure drop favours an alternative exfoliation pathway. To do this we have developed a simple model to calculate the pressure required to peel off a single layer flake from its bulk counterpart. For the sake of definiteness we consider graphite. Other layered materials can be treated analogously. A graphene sheet at the surface of a crystallite experiences an attractive force towards the neighbouring sheet that can be derived from the Lennard-Jones potential[101,102]

$$U(x) = 4\,A\,\varepsilon_0\left[\left(\frac{\sigma}{x}\right)^{12} - \left(\frac{\sigma}{x}\right)^{6}\right],$$

where $A$ is the sheet area and $x$ is the distance between two neighbouring sheets. The equilibrium distance is $x_0 = 2^{1/6}\sigma$ with $\varepsilon_G = -U(x_0)/A$ the energy per unit area necessary to complete the expansion process,[101] if the free energy of the solvent is neglected. Rigid oscillations of the graphene sheet take place around the equilibrium distance, with angular frequency $\Omega = \sqrt{36\,\varepsilon_G\,A_0/(m_C x_0^2)}$, where $m_C$ is the carbon mass and $A_0$ is the area of the primitive graphene cell in real space. To evaluate the impact of the pressure drop across the nozzle on the graphitic particles, we parameterize it with an exponential profile $P(t) = P_f + e^{-t/\tau}(P_i - P_f)$, where the transit time $\tau$ is related to the nozzle length $L$ and the flow speed $v$ by $\tau = L/v$ and $P_i$ ($P_f$) is the pressure before (after) the nozzle. We then solve the equation of motion for the harmonic oscillations of the distance $x$ in the Lennard-Jones potential, in the presence of the force due to the pressure:

$$\ddot{x}(t) = -\Omega^2[x(t) - x_0] - \frac{A_0}{2m_C}P(t),$$

with the initial condition $x_i = x_0 - A_0 P_i/(2\,m_C \Omega^2)$ found by requiring that the total force vanishes. The initial potential energy $U(x_i)$ is larger than $U(x_0)$, because of the work performed by the pressure force. As the pressure drops and the oscillator relaxes from its initial position, work is dissipated by the oscillating sheet into the solvent exerting the pressure. The total dissipated work is:

$$W = \int dW = \int dx\,A\,P = \int dt\,A\,P\,\dot{x}.$$

Using that the frequency of the oscillations is much larger than $1/\tau$, we obtain the following compact expression for the work dissipated per unit area

$$\frac{W}{A} = \frac{A_0}{4m_C}\frac{\tau^2}{1 + (\Omega\tau)^2}.$$

Expansion is activated if the initial potential energy minus the work dissipated into the fluid is larger than the potential energy of the sheet at large $x$, which vanishes, i.e., $U(x_i) - W > 0$.

This condition can be conveniently rewritten into a condition on the rate of pressure drop in time as follows:

$$\dot{P} \sim \frac{P_i}{\tau} > \Omega^2 \sqrt{\frac{4\,m_C}{A_0}} \sqrt{\varepsilon_G}.$$

In deriving this equation, we have used for simplicity that $P_i \gg P_f$ and $\Omega\tau \gg 1$. Using $m_C$ = 20.4 $10^{-27}$ kg, $A_0$ = 0.051 nm², $\varepsilon_G$ = 71 mJ m⁻², and $x_0$ = 0.34 nm, we find $\Omega/2\pi$ ~ 1.2 THz, which is of the same order of magnitude of the ZO' vibration mode in FLG.[103,104]

The flow speed $v$, estimated using the nozzle diameter and the dispersion flux, is of the order of $v$ ~ $10^3$ km h⁻¹. Given the length of the nozzle $L$ = 0.95 mm, we have $\tau$ ~ 3.4 μs, which finally leads to $P_i$ > 6x10¹⁶ Pa. This enormous value shows that expansion, *i.e.*, separation of two adjacent crystal sheets in the normal direction, is not active in the WJM (which reaches a maximum pressure of 250 x10⁶ Pa) because it requires much larger pressure drops to be activated. Therefore, we conclude that exfoliation in the WJM is dictated by shear forces.

## Experimental part

### Exfoliation of Graphite in N-methyl-2-pyrrolidone (NMP)

A mixture of the bulk layered crystals (200 g of graphite flakes +100 mesh from Sigma Aldrich) and the solvent (20 L of NMP, Sigma Aldrich) is prepared. The mixture is placed in the container and mixed with a mechanical stirrer (Eurostar digital Ika-Werke). For the +100 mesh graphite the 0.30 mm nozzle aperture is used. The piston-pass, defined as the number of times the piston is charged and discharged with solvent/layered crystal, is set to 1000 passes (10 mL per pass).[91] The processed sample, named WJM0.30, is then collected in a second container. The wet-jet milling process is repeated passing the sample WJM0.30 through the 0.15 mm nozzle. The corresponding processed sample is named WJM0.15. Finally, the nozzle is changed to 0.10 mm diameter and a third exfoliation step is carried out. This sample is named WJM0.10.

### Exfoliation of other layered crystals

We select $h$-BN (~1 μm, 98%, from Sigma Aldrich), WS₂ (2 μm, 99%, Sigma Aldrich) and MoS₂ (<2 μm, 99%, Sigma Aldrich). Since the crystallite size of these materials is much smaller than the typical lateral size of graphite flakes (~150 μm), it is possible to perform the exfoliation directly with the 0.10 mm nozzle diameter. 100 g of each material is dispersed in 10 L of NMP.

### Purification of the WJM-treated sample

For high quality flakes with defined lateral size and thickness, a post-processing procedure is required to purify/separate single- and few-layer 2D crystals from the thicker ones (>10 layers). For this purpose, we use the sedimentation based separation (SBS). The SBS is usually applied to particles[105,106] or flakes[71,107] dispersed in a solvent under a force field.[105] The forces acting in the SBS are the centrifugal force $F_c = m_p \omega^2 r$, proportional to the mass of the particle itself ($m_p$), the distance from the rotational axes ($r$), the square of the angular velocity ($\omega$), the buoyant force $F_b = -m_s \omega^2 r$, which is equal to the mass of the displaced solvent ($m_s$) times the centrifugal acceleration, and the frictional force $F_f = -f\sigma$, i.e., the force acting on the particles while moving with a sedimentation velocity ($\sigma$) in a fluid.[105] This force is proportional to the friction coefficient ($f$) between the solvent and the particle itself. The sum of the forces acting on the dispersed flakes is $F_c - F_b - F_f = F_{tot}$.[96] Defining the sedimentation coefficient ($S$) as the ratio between the $v$ and the centrifugal acceleration, we can write:

$$S = \frac{\sigma}{\omega^2 r} = \frac{m_p \left(1 - \frac{\rho_s}{\rho_p}\right)}{f}$$

where $\rho_s$ and $\rho_p$ are the density of the solvent and the particle, respectively (see Refs. 86 and 108 for details concerning its derivation). Thick and large flakes sediment faster than thin and small flakes due to larger sedimentation coefficients compared with the small flakes.[108] By tuning the experimental centrifugation parameters, it is possible to retain flakes with different lateral sizes in dispersion. (See Figure 2.) For graphene purification, we performed the SBS at two centrifugal accelerations, i.e. ~500 $g$ (gravitational acceleration corresponding to 2000 rpm for the rotor used) and ~3000 $g$ (5000 rpm) for 30 min. For the purification of the exfoliated $h$-BN, MoS$_2$ and WS$_2$ flakes, we performed the SBS at ~3000 $g$ (5000 rpm) for 30 min. The centrifugations are carried out in a Coulter-Beckman Ultracentrifuge Optima XE-90, using a SW32Ti rotor. After the centrifugation, the upper 80% of the supernatant is taken, discarding the precipitate.

## Optical extinction spectroscopy (OES)

The samples are prepared by diluting the dispersions of all the prepared 2D-materials in NMP with ratio 1:50. OES is carried out in a Cary Varian 5000UV-Vis spectrophotometer. The concentrations, $C$, are determined from the optical absorption coefficient at 660 nm, using the Lambert law $A = \alpha l c$ where $l$ [m] is the light path length, $c$ [gL$^{-1}$] is the $C$ of dispersed material, and $\alpha$ [Lg$^{-1}$m$^{-1}$] is the extinction coefficient. The extinction coefficient used for graphitic flakes is $\alpha$ ~1390 Lg$^{-1}$m$^{-1}$ at 660 nm.[109] The extinction coefficients used to calculate the $C$ of boron nitride, WS$_2$ and MoS$_2$ dispersions are $\alpha_{300nm}$ ~2367 Lg$^{-1}$m$^{-1}$, $\alpha_{629nm}$ ~2756 Lg$^{-1}$m$^{-1}$, and $\alpha_{672nm}$ ~3400 Lg$^{-1}$m$^{-1}$, respectively, where the subscript is the wavelength used for the measurement.[72]

## Transmission electron microscopy (TEM)

Graphene, $h$-BN, MoS$_2$ and WS$_2$ are prepared by drop casting dispersions onto ultrathin C-film on holey carbon 400 mesh Cu grids, from Ted Pella Inc. The graphene samples are diluted 1:50, while the h-BN, MoS$_2$ and WS$_2$ samples are diluted 1:20. The grids are stored under vacuum at room temperature to remove the solvent residues. TEM images are taken by a JEOL JEM-1011 transmission electron microscope, operated at an acceleration voltage of 100 kV. High-resolution TEM (HRTEM) is performed using a 200 kV field emission gun, a CEOS spherical aberration corrector for the objective lens, enabling a spatial resolution of 0.9 Å, and an in-column image filter (Ω-type).

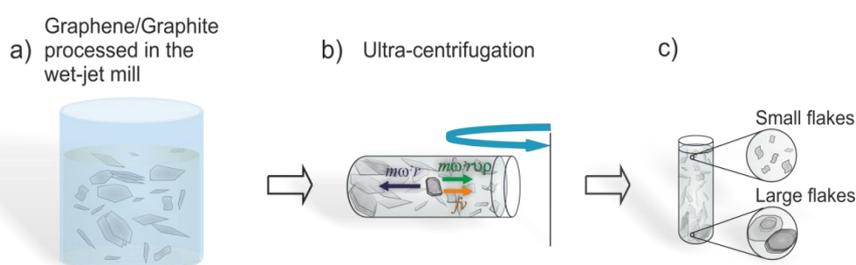

**Figure 2.** Purification of the wet-jet milled graphite (a) done by sedimentation based separation and after applying a centrifugal force (b), where the flakes arrange according to their densities. In this way, (c) the less thick and smaller flakes are in the upper part of the centrifuge tube and the large or un-exfoliated flakes are at the bottom of the centrifuge tube.

## Raman spectroscopy

The as-prepared dispersions are diluted 1:30 in NMP and drop-cast onto a Si wafer (LDB Technologies Ltd.) covered with 300 nm thermally grown SiO$_2$. The bulk materials are analysed in the powder form. Raman measurements are carried out by a Renishaw inVia spectrometer using a 50× objective (numerical aperture 0.75), a laser with a wavelength of 514.5 nm with an incident power of ~5 mW. A total of 30 points per sample are measured to perform the statistical analysis. OriginPro 2016 is used to perform the deconvolution and statistics.

## Atomic Force Microscopy (AFM)

The dispersions are diluted 1:30 in NMP. 100 µL of the dilutions are drop-cast onto Si/SiO$_2$ wafers, and dried at 50°C overnight. AFM images are acquired with a Bruker Innova® AFM in tapping mode using silicon probes (frequency = 300 kHz, spring constant = 40 Nm$^{-1}$). Thickness statistics is performed by measuring ~100 flakes from the AFM images. Statistical analyses are fitted with log-normal distributions. Statistical analysis are performed in WSxM Beta 4.0 software.[110]

### X-ray photoelectron spectroscopy (XPS)

The analysis is accomplished using a Kratos Axis UltraDLD spectrometer on samples drop-cast onto gold-coated silicon wafers. The XPS spectra are acquired using a monochromatic Al K$_\alpha$ source operating at 20 mA and 15 kV. The analyses are carried out on a 300 μm × 700 μm area. High-resolution spectra of C 1s and Au 4f peaks were collected at pass energy of 10 eV and energy step of 0.1 eV. Energy calibration is performed setting the Au $4f_{7/2}$ peak at 84.0 eV. Data analysis is carried out with CasaXPS software (version 2.3.17).

### Polyamide–12-graphene composite

The composite is prepared by melt blending. The as-produced WJM0.1 sample is dried using a rotary evaporator (Heidolph, Hei-Vap Value, at 70 °C, 5 mbar). Polyamide–12 (Sigma Aldrich) and the graphene WJM0.10 powder (1% in weight) are mixed in a twin-screw extruder (model: 2C12-45L/D, Bandiera) at 175°C. The mechanical properties of bare Polyamide–12 and Polyamide–12/graphene composites are measured using a universal testing equipment (Instron Dual Column Tabletop System 3365), with 5 mm/min cross-head speed. The tensile strength measurements are carried out on 7 different samples for each composite material according to ASTM D 882 Standard test methods.

### Li-ion battery anodes fabrication

Round-shape Cu disks (diameter of 1.5 cm, thickness of 25 μm, Sigma-Aldrich) are cleaned with acetone and 2-propanol (Sigma-Aldrich) in ultrasonic bath for 10 min. Then, the Cu disks are dried and weighted (Mettler Toledo XSE104). Subsequently, 100 μL of as-prepared WJM0.10 are drop-cast on each Cu disk under air atmosphere at 80 °C and then dried at 120 °C and $10^{-3}$ bar for 12 hours in a glass oven (BÜCHI, B-585). The graphene mass loading (~1 mg) for each electrode is calculated by subtracting the weight of bare Cu foil from the total weight of the electrode.

### Half-cells assembling and electrochemical characterization

The graphene-based electrodes (anode) are tested against circular Li foil (Sigma-Aldrich) in the half-cell configuration, and assembled in coin cells (2032, MTI) in an argon-filled glove box ($O_2$ and $H_2O$<0.1 ppm) at 25 °C, using 1 M $LiPF_6$ in a mixed solvent of ethylene carbonate/dimethylcarbonate (EC/DMC, 1:1 volume ratio) as electrolyte (LP30, BASF) and a glass fibre separator (Whatman GF/D). The cyclic voltammetries (CVs) are performed at a scan rate of 50 μVs$^{-1}$ between 1 V and 5 mV vs Li$^+$/Li with a Biologic, MPG2 potentiostat/galvanostat. The constant current charge/discharge galvanostatic cycles are performed for the as-prepared graphene based anodes in half-cell configurations using a battery analyser (MTI, BST8-WA). All the electrochemical measurements are performed at room temperature.

## Ink-jet printing

WJM0.10 viscosity is measured by a Discovery HR-2 Hybrid Rheometer (TA instruments), using a double-wall concentric cylinders geometry (inner diameter 32 mm, outer diameter 35 mm), designed for low-viscosity fluids. The temperatures of the dispersions are set and maintained at 25 °C throughout all the measurements. 2 mL of WJM0.10 are then loaded into a cartridge reservoir (fluid bag, Fujifim Dimatrix, DMC-11610). The WJM0.10 is ink-jet printed on $SiO_2$/Si by a Fujifilm Dimatix 2800 printer. The printed pattern is then annealed at 1000 °C for 1 hour under a $H_2$/Ar gas atmosphere. The sheet resistance of the printed pattern is measured with a four-point-probe test unit (Jandel, model RM300). Film thickness is determined by a Mitutoyo Absolute Digimatic Micrometer Series 227 with a pressure of 1N, 1.25 µm resolution.

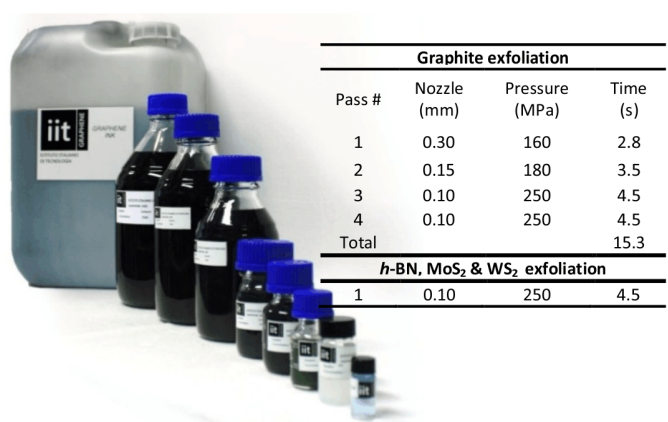

| Pass # | Nozzle (mm) | Pressure (MPa) | Time (s) |
|---|---|---|---|
| Graphite exfoliation | | | |
| 1 | 0.30 | 160 | 2.8 |
| 2 | 0.15 | 180 | 3.5 |
| 3 | 0.10 | 250 | 4.5 |
| 4 | 0.10 | 250 | 4.5 |
| Total | | | 15.3 |
| $h$-BN, $MoS_2$ & $WS_2$ exfoliation | | | |
| 1 | 0.10 | 250 | 4.5 |

**Figure 3.** Graphene–NMP, $WS_2$-NMP and $h$-BN-NMP dispersions produced by Wet-jet mill. The inset table shows the pressure and time required to process 10 mL of layered crystals-NMP.

## Results and discussion

In the case of 100+ mesh graphite, we chose the 0.3 mm nozzle diameter to start the exfoliation process. We experimentally observe that starting the exfoliation of such large crystallites with smaller nozzles (<0.3 mm) may cause system clogging. According to this consideration, several combinations with different nozzle diameters (0.30, 0.20, 0.15 and 0.10 mm) are tested. Graphite exfoliation is achieved using the following nozzle diameters sequence: 0.30 mm, 0.15 mm, and twice 0.10 mm. Considering all the four passes, the effective time required to process 10 mL of graphite/NMP ($C$ of 10 gL$^{-1}$), is 15.3 s. The processing time for each nozzle is reported in the inset to Figure 3 (table).

In the case of $MoS_2$, $WS_2$, and $h$-BN, their crystallite sizes allow to use the 0.10 mm nozzle directly, giving a processing time of 4.5 s per 10 mL of sample. Figure 3 shows the production samples up to 30 L of graphene, 50 mL of $WS_2$, and 20 mL of $h$-BN dispersions, produced by the WJM technique.

After exfoliation by WJM of the layered crystals, TEM and AFM are performed to analyse flake sizes and thickness, respectively, of the as-prepared 2D crystals. For what concerns the exfoliation of graphite, TEM analysis indicates that the flake main lateral size distribution decreases from 149 μm (starting graphite material) to 1000 nm (Log-Normal standard deviation, SD: 0.53), 850 nm (SD: 0.83), and 460 nm (SD: 1.18) for WJM0.30, WJM0.15, and WJM0.10, respectively (see Figure 4a, b, and c). The lateral-size distribution statistics are reported in the inset to Figure 4a-c, obeying a log-normal distribution, which is the typical distribution for fragmented systems[111] (See the Appendix for the discussion of the nomenclature used in this paper). The number of layers in the WJM0.10 sample can be directly visualized on HRTEM in a bended-flake edge.[112] Figure 4d shows a representative bended flake with three layers, demonstrating that the WJM0.10 is composed by few-layer graphene flakes. At higher magnification (Figure 4e) the honeycomb carbon lattice can be observed.[113] The upper inset shows the lattice parameter 0.247 nm of graphite (hexagonal, $p6_3/mmc$ #194, a=b=2.4 Å, c=6.70 Å)[113] with the indexed fast Fourier transform given in the lower inset.

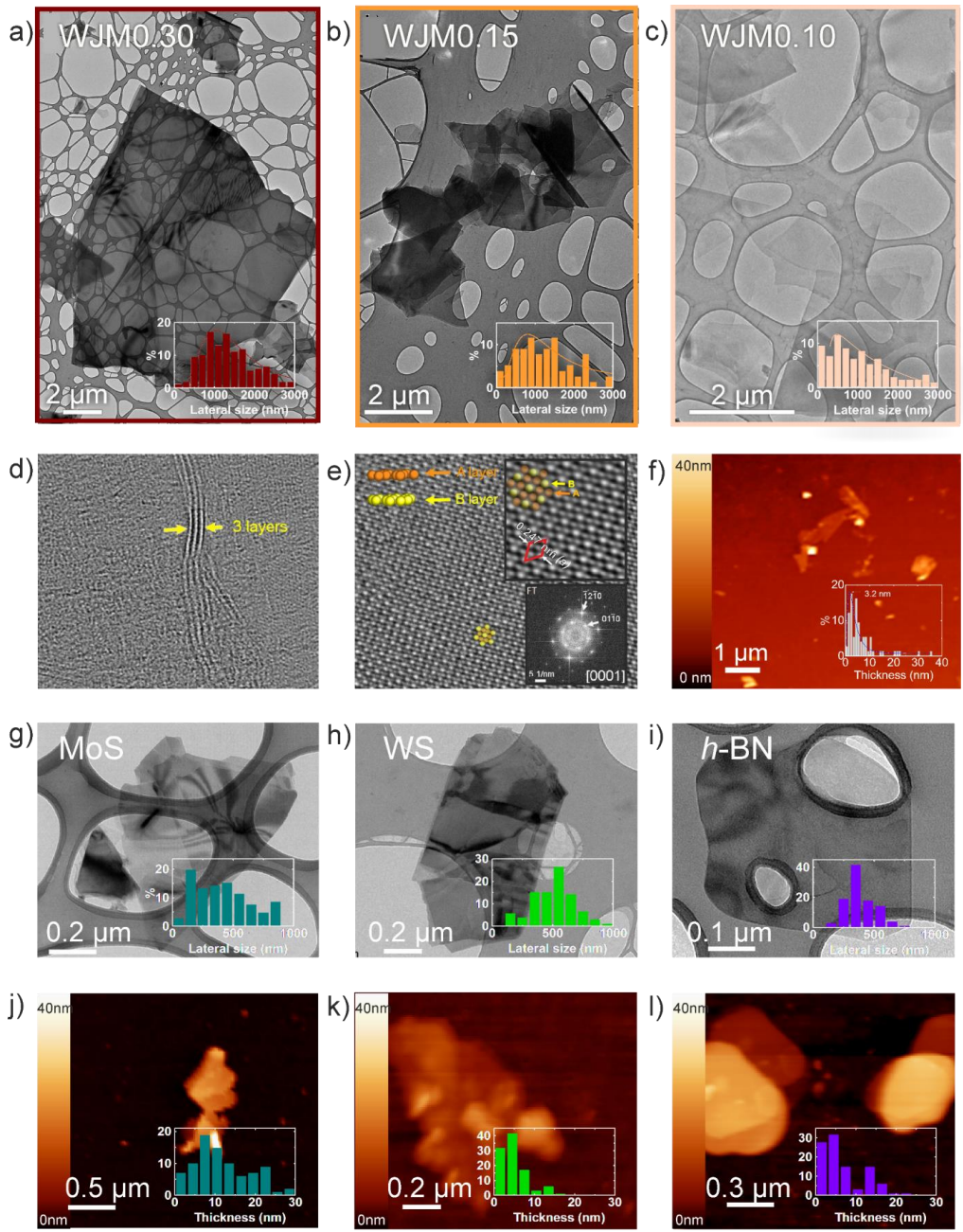

**Figure 4**. (a, b, and c) TEM images of WJM0.10, WJM0.15 and WJM0.30 samples, respectively, with the corresponding lateral size distributions in the insets, displaying a Log-normal distribution. (d) HRTEM image of a bended three-layer graphene, from sample WJM0.10. (e) HRTEM image of a WJM0.10 flake showing the A-B stacking, depicted by orange and yellow circles on the image. Inset in the upper right corner reports a zoomed area on the same flake where the red diamond indicates the hexagonal unit cell of graphite. The bottom right inset shows the corresponding Fourier transform with the indexed reflections from crystalline planes. (f) AFM image of graphene flakes deposited on Si/SiO$_2$ substrate with the thickness distribution given in the inset. The maximum population of flake thickness peaks at 3.2 nm. (g, h, and i) TEM images of exfoliated MoS$_2$, WS$_2$, and

*h*-BN flakes, respectively, and their corresponding statistical lateral size distributions are shown in the insets. (j, k, and l) AFM images of exfoliated flakes of $MoS_2$, $WS_2$, and *h*-BN flakes respectively. Their corresponding thickness distributions are shown in the insets.

The thickness of the exfoliated flakes is analysed by AFM. Figure 4f shows the AFM image of an exfoliated sample WJM0.10, giving a main thickness distribution of 3.2 nm (see inset to Figure 4f).

The $MoS_2$, $WS_2$, and *h*-BN flakes sizes are also determined by TEM, with the images of the exfoliated flakes shown in Figure 4g-i. The lateral-size statistical distribution of the flakes, shown in the insets, displays an average size of 380, 500, and 340 nm for $MoS_2$, $WS_2$, and *h*-BN, respectively. The thickness of the processed crystals is analysed by AFM (Figure 4j to l). The insets on each image report the statistical distribution of the thicknesses, showing a thickness mode at 6, 4.5, and 2.4 nm for $MoS_2$, $WS_2$, and *h*-BN, respectively.

The quality of the exfoliated material, in terms of crystalline integrity, is analysed by Raman spectroscopy. The Raman spectra of graphene consist mainly on the D, G, and 2D band (the latter composed by $2D_1$ and $2D_2$ contributions, see the Appendix for information regarding the labelling of the peaks). For graphene obtained by LPE, it is uncommon to find such large intensities for the 2D band, even for SLG. [56,63,109,114] Taking into account the intensity ratios of the $2D_1$ and $2D_2$ bands (see Figure 5a), it is possible to estimate the flake thickness. Figure 5a shows the Raman spectra of the samples WJM0.30, WJM0.15, WJM0.10, and graphite for comparison. All spectra are normalised to the G peak intensity.

The intensity variations of the D and D' bands are related to an increase of edge or in-plane defects.[115,116,117] The statistical analysis shows that I(D)/I(G) ranges from 0.03 to 0.6 for WJM0.30, then the range varies to 0.1 - 1 and 0.1 - 1.2, for WJM0.15 and WJM0.10 samples, respectively (Figure 5b). On the contrary, FWHM(G) (Figs 5c) are not significantly affected by the nozzle diameter, ranging in all cases from ~14 to ~25 cm$^{-1}$, with a mode at ~19 cm$^{-1}$. The plot of I(D)/I(G) vs. FWHM(G) shows that the linear correlation between these parameters becomes scattered (not correlated) when the nozzle diameter is reduced. In fact, the linear correlation is also reduced from 0.748 to 0.289 for WJM0.30 to WJM0.10. This result suggests that the WJM process homogenises the sample by increasing the quantity of flakes smaller than the laser spot size (1μm).[150] This is in agreement with the TEM measurements. The normalised intensity ratios $I(2D_1)/I(G)$ vs. $I(2D_2)/I(G)$ give an insight on the flake thickness (see Figure 5e). In general, for graphite, the intensity of $2D_2$ peak [$I(2D_2)$] is roughly double compared to the intensity of $2D_1$ peak [$I(2D_1)$][118].

Furthermore, the intensity ratio [$I(2D_2)/I(2D_1)$] decreases as the flake thickness is reduced,[119] until the 2D band can be fitted by a single Lorentzian, highlighting that the flakes are electronically decoupled.

The dashed line in Figure 5e represents the multilayer condition (~5 layers)[120,121] [I(2D$_1$)/I(G) = I(2D$_2$)/I(G)] separating the data set, while the points below the line [I(2D$_1$)/I(G) < I(2D$_2$)/I(G)] are considered graphitic flakes, and the points above the line [I(2D$_1$)/I(G) > I(2D$_2$)/I(G)] are considered FLG and SLG.[120,121] It is noteworthy that a single Lorentzian component is achieved only for the sample WJM0.10, indicating that the graphite processing through the 0.10 mm nozzle allows to get graphene flakes with electronically decoupled layers. Additionally, the evolution of the 2D band when graphite is processed through the nozzles indicates an effective reduction of flake thickness. Figure 5f presents the high-resolution XPS C 1s spectrum of WJM0.10. The spectrum can be decomposed into different components typical of graphite: a main peak at 284.4 eV for sp$^2$ carbon with the corresponding feature due to π-π* interactions at 290.8 eV, as well as a second peak at 284.8 eV for sp$^3$ carbon, due to flake edges and solvent residual. The sp$^3$ fraction is around 26%.

Two other weak contributions equal to ~10% of the total carbon amount can be ascribed to C-N and C=O groups (peaks at binding energies = 286.3 eV and 287.7 eV). These nitrogen and oxygen groups likely come from residual NMP molecules.

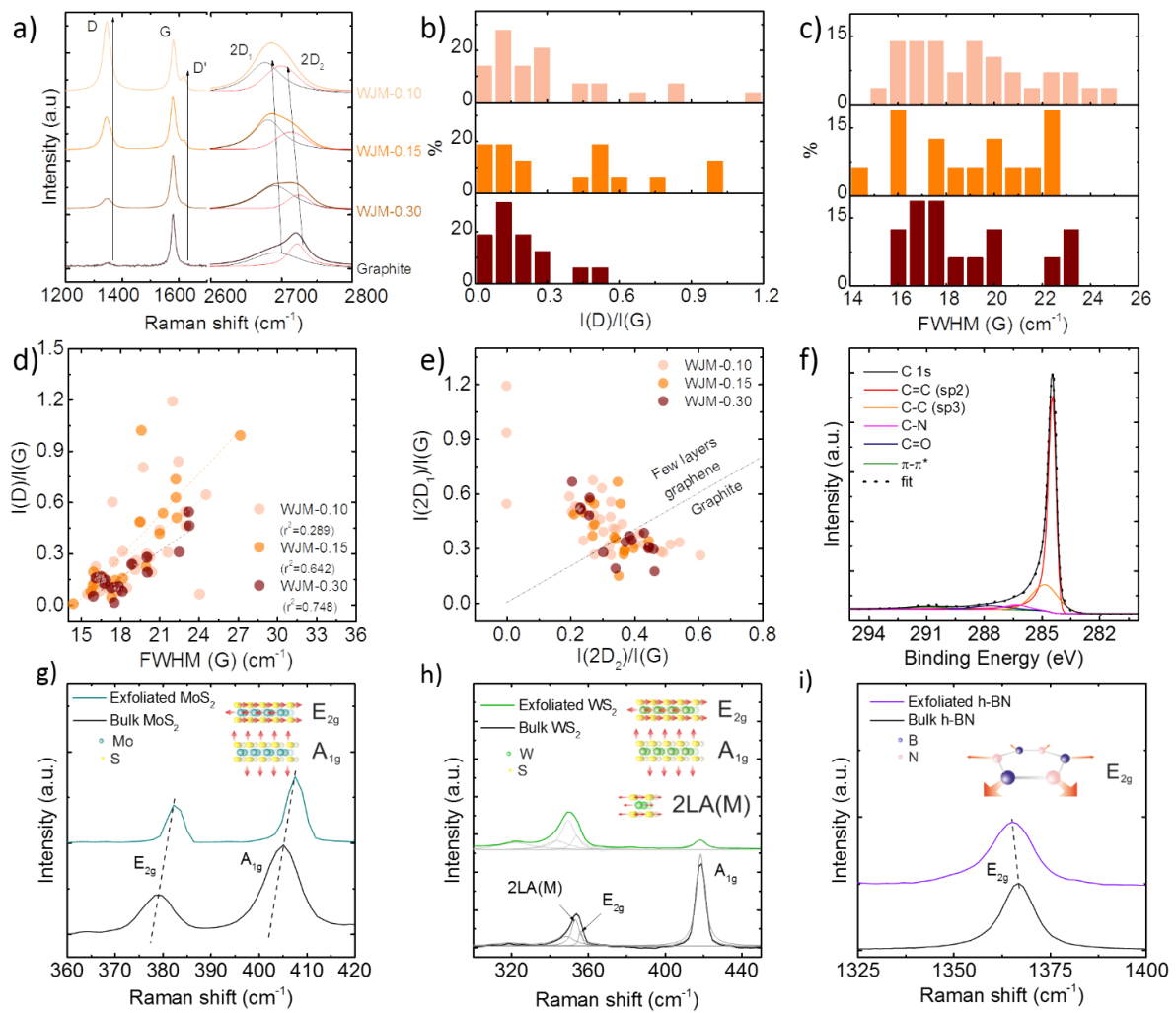

**Figure 5.** (a) Raman spectra of the samples WJM0.10, WJM0.15, and WJM0.30 in pink, orange, and wine, respectively, and graphite spectrum (in black), for the sake of comparison. The black arrows indicate the increase in the integral intensity of the D and D' peaks, and a shift for the $2D_1$ and $2D_2$ peaks when the samples are processed from graphite to WJM0.10. The statistical analyses of the I(D)/I(G) and FWHM(G) of the processed samples are shown in (b) and (c). (d) FWHM(G) vs. I(D)/I(G) and their linear correlation (dashed line) and (e) the normalised integral intensities of the peaks $2D_1$ and $2D_2$ showing the distribution of FLG and graphite. The dashed line represents the condition where $I(2D_1)/I(G) = I(2D_2)/I(G)$. (f) XPS C 1s spectrum of WJM010. (g, h, and i) Raman spectra of the 2D crystals compared with their bulk counterparts. The Raman active modes are illustrated as insets in each figure.

Raman spectroscopy is also used to analyse the physical changes on the exfoliated $MoS_2$, $WS_2$, and $h$-BN samples. The Raman spectra of bulk and exfoliated $MoS_2$, $WS_2$, and $h$-BN are shown in Figures 5g-i, and their corresponding vibrational modes are illustrated in the insets to the corresponding figures. The Raman spectrum of bulk $MoS_2$ consists of two active peaks, the first one ($E_{2g}$), at 379 cm$^{-1}$, corresponds to the mode involving the in-plane vibration of Mo and S atoms.[122,123] The second one ($A_{1g}$), at 405 cm$^{-1}$, is due to out-of-plane vibrations.[122,123] The typical Raman spectra of exfoliated $MoS_2$ show a shift of the $E_{2g}$ and $A_{1g}$ peaks, such that the distance between the peaks goes from 26 cm$^{-1}$ for the bulk case to 19 cm$^{-1}$ in the monolayer limit.[123,125,126] The $MoS_2$ Raman spectrum of the exfoliated samples is reported in Figure 5g, blue line. The spectrum shows a blue shift for both bands, $E_{2g}$ (3 cm$^{-1}$) and $A_{1g}$ (4 cm$^{-1}$), with respect to the bulk case. Similar results have been reported for exfoliated $MoS_2$ flakes.[124] The Raman spectrum of $WS_2$ consists mainly of three peaks: the $E_{2g}$, which corresponds to the mode involving the in-plane vibration of W and S atoms; the $A_{1g}$, which is related to out-of-plane vibrations; and the second-order longitudinal acoustic mode (2LA) at 350 cm$^{-1}$.[125,126,127] The integral intensity of the 2LA peak increases with the decreasing flake thickness.[125,126,127]

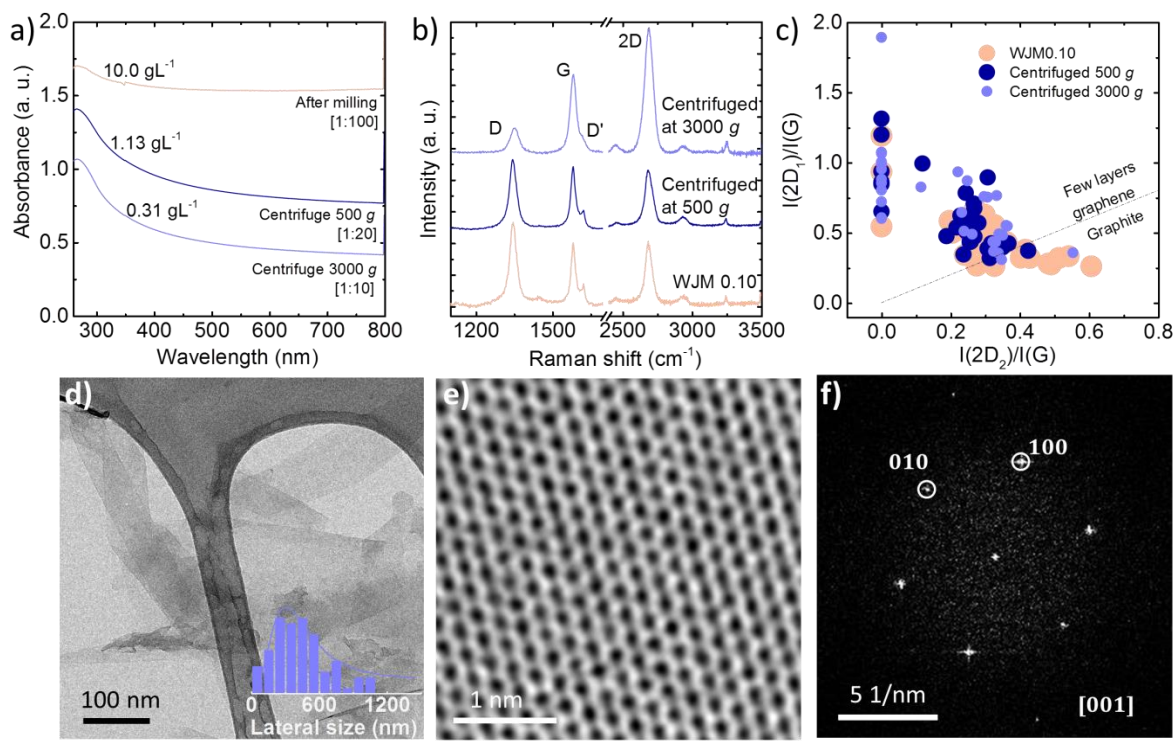

Figure 6. (a) Optical absorption spectroscopy and (b) Raman spectra of the sample WJM0.10 and purified after centrifugation at 500 $g$ and 3000 $g$. (c) The normalised integral intensities of the $2D_1$ and $2D_2$ peaks show the distribution of FLG and graphite in the purified sample at 3000 $g$ and, as comparison, the as-produced WJM0.10 sample. (d) TEM image of the centrifuged sample at 3000 $g$, the inset shows the lateral size distribution. (e) HRTEM image of a flake reported in (d). (f) Fast Fourier Transform of the flake in (e) with the indexed reflections from crystalline planes.

The spectrum of exfoliated $WS_2$ (Figure 5h, green line) shows a 7-fold decrease of the integral intensity of the $A_{1g}$ mode and a two-fold increase in the intensity of the 2LA phonon mode, due to the occurrence of a double resonance for exfoliated $WS_2$ flakes, , in agreement with previous studies related with the exfoliation of $WS_2$.[125,126,127] Lastly, the bulk $h$-BN Raman spectrum exhibits a single peak located at 1366 cm$^{-1}$ ($E_{2g}$), which is due to in-plane atomic displacements.[128] The Raman spectrum of exfoliated $h$-BN (Figure 5i, purple line) shows a broadening of the $E_{2g}$ band, characteristic of exfoliated $h$-BN samples.[129] Additionally, a shift of the $E_{2g}$ band has been explained as a result of stress induced in the exfoliation process.[130] Detailed information on the Raman spectra and statistics can be found in the Appendix. In summary, the TEM, AFM, and Raman results demonstrate successful exfoliation of the layered crystals. The as-produced exfoliated samples consist of a mixture of flakes of different thicknesses, as discussed above. The thick flakes in the sample can be removed by SBS, thus promoting sample enrichment with thin flakes, as described in the Methods section. The initial $C$ of the sample WJM0.10 is confirmed by OES to be~10 gL$^{-1}$ (see

Figure 6a, orange line). After centrifugation, the value of C of flakes in dispersion decreases to 1.13 gL$^{-1}$ and 0.31 gL$^{-1}$ for 500 g (dark blue line) and 3000 g (light blue line), respectively (Figure 6a). The physical changes of the WJM0.10 samples after centrifugation are also evident in Raman spectroscopy (Figure 6b), from the changes of the normalised intensity ratios I(2D$_1$)/I(G) vs. I(2D$_2$)/I(G) band (Figure 6c). For the WJM0.10 sample, the points satisfying the condition I(2D$_1$)/I(G) < I(2D$_2$)/I(G) (>5 layers[120,121]) decrease from 37% to 7%, for the centrifuged samples. Conversely, the points having I(2D$_2$)/I(G) ≈ 0 increase from 10%, for the as-produced WJM0.10 to 40% for the sample centrifuged at 500 g, and to 57% for the sample centrifuged at 3000 g. These results indicate that the SBS is an effective process to separate graphite-like flakes from FLG.[131]

Finally, in order to gain further insight on the quality of the purified samples, we additionally analysed the sample centrifuged at 3000 g by TEM and HRTEM (Figure 6d-f). The statistical lateral size distribution, shown in the inset to Figure 6d, peaks at 350 nm. It is worth noting that the log-normal standard deviation decreases from 1.18 for WJM0.10 to 0.55, meaning that the centrifuged sample has a narrower lateral size distribution than the WJM0.10. The HRTEM image (Figure 6e) of one of the flakes shows the characteristic honeycomb lattice of graphene. The corresponding Fast Fourier Transform (Figure 6f) suggests the absence of multi-layered structures or stacked flakes. In summary, these results indicate that the WJM is an ideal tool to produce gram-scale quantities of FLG flakes, and also SLG with the usage of purification procedures.

## Applications of graphene obtained by wet-jet mill

The graphene flakes produced by WJM can be used in applications wherein large quantities of high-quality flakes are required. As a proof of concept, we select three applications in which graphene obtained by the WJM (*i.e.*, sample WJM0.10) improves the applications' performance, demonstrating the WJM as a promising process for the industrial exploitation of exfoliated layered crystals.

### WJM0.10 as anode material for lithium ion batteries (LIBs)

The current commercial graphite anodes of Li-ion batteries (LIBs) have a theoretical specific capacity limited to 372 mAh g$^{-1}$.[132,133] SLG/FLG are possible candidates to replace graphite as active anode material and improve the performance of LIBs[17,18] although issues related to irreversible processes induced by the large surface exposed to electrolytes should be considered.[134] For this purpose, the as-prepared WJM0.10 SLG/FLG samples is tested as anode-active-material for LIBs. The voltage profiles of WJM0.10 as anode-active material, obtained by electrochemical tests in half cell configuration against Li foil, show that >85% of the capacity is delivered at a potential lower than 0.25 V vs. Li$^+$/Li, with a flat plateau up to the 20$^{th}$ cycle (Figure 7a). The working voltage is comparable to the values obtained using commercial graphite anodes (0-0.4 V vs. Li$^+$/Li), leading to a high-energy efficiency of batteries.[2] An irreversible capacity of 100 mAh g$^{-1}$ is observed during the 1$^{st}$ charge/discharge cycle. Additionally, the WJM0.10-based anode gives a specific capacity of ~420 mAh g$^{-1}$ at current density of 0.1 A g$^{-1}$ after 50 charge/discharge cycles and a Coulombic efficiency (the discharge capacity vs. charge capacity) of 99.8% (Figure 7b).

Compared to other graphene-related materials, including graphene oxide,[135,136] reduced graphene oxide,[137] or pristine graphene,[17,18,19] which deliver much higher irreversible capacities (200-5000 mAh g$^{-1}$),[135,136,138] WJM0.10 results to be a promising candidate as anode material for LIBs. In fact, firstly, WJM0.10 is used as-prepared after the WJM process, *i.e.*, without purification. Secondly, it shows low working voltage (0.25 V vs Li$^+$/Li), small irreversible capacity (100 mAh g$^{-1}$), and high Coulombic efficiency (99.8%).

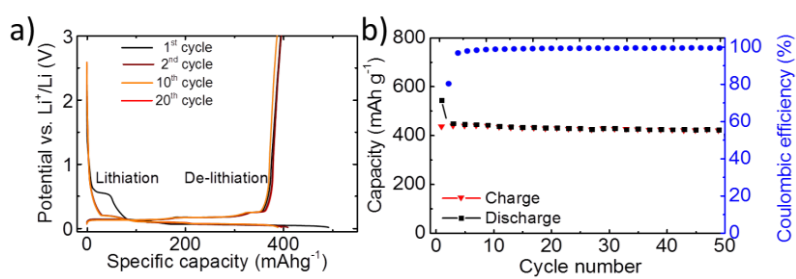

**Figure 7** (a) Voltage profile upon galvanostatic charge/discharge of graphene anode at 0.1 A g$^{-1}$ between 50 mV and 3 V. (b) Specific capacity and coulombic efficiency over galvanostatic cycles at current density of 0.1 A g$^{-1}$ between 50 mV and 2 V.

### WJM0.10 as reinforcement of Polyamide–12

One of the most feasible applications of graphene is as filler in polymer matrices improving the physical properties of the matrix.[139,140,] In recent studies, we demonstrated that graphene flakes produced by LPE improve the tensile modulus of polycarbonate-based composite (+26% enhancement of the elastic modulus at 1 w%),[87] and the size of the flakes (thickness and lateral size) influence the mechanical properties of graphene/polymer composites.[12]

In this regard, the amide-based polymers (typical engineering thermoplastic materials)[141] are commonly used for electric, food, and pharmaceutical packaging, and the improvement of the mechanical properties (*e.g.,* strength) of polyamide is relevant for the packaging industry.[142] In this regard, we test the as-produced WJM0.10 as mechanical reinforcement of Polyamide–12 (PA12), see Experimental part. The WJM0.10 flakes in the PA12 matrix are shown in the false-coloured SEM image (Figure 8a). We measured the flexural modulus, defined as the slope of the flexural stress vs.

flexural strain curve in the elastic region,[141] in order to evaluate the mechanical improvement in the composite. Representative flexural stress vs. flexural strain curves of pristine PA12 (black curve) and PA12/WJM0.10 (blue curve) loadings are shown in Figure 9b. The flexural modulus increases from 1412 MPa (bare PA12) to 1890 MPa (composite), corresponding to a 34% improvement. This PA12/WJM0.10 composite could be exploited, as an example, in packaging-related applications, in which the integrity of the material must be guaranteed under deformation.[143]

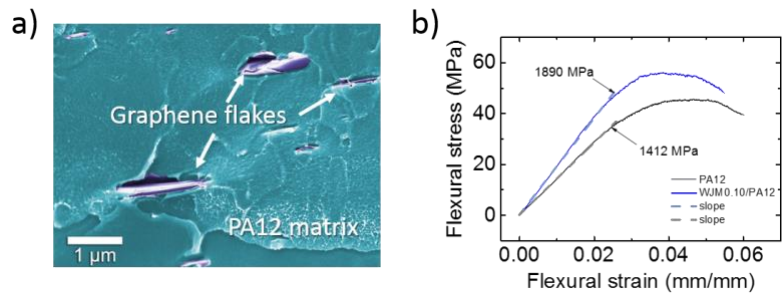

**Figure 8.** (a) False-coloured scanning electron micrograph of PA12/WJM0.10 composite. (b) Flexural modulus (stress vs. strain) of PA12 and PA12/WJM0.10 composites with 0.5% in weight of graphene loading.

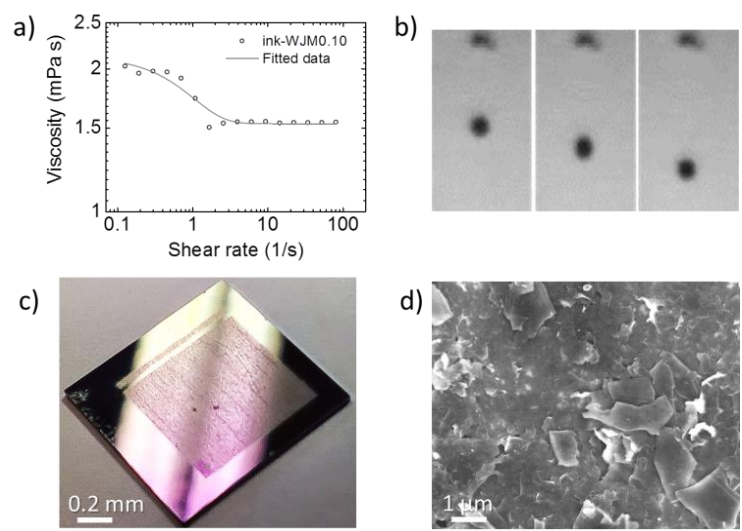

**Figure 9.** (a) Viscosity vs. shear rate of the ink (WJM0.10). (b) Jetting from the cartridge nozzle of the ink (WJM0.10). (c) Image of printed paths. (d) SEM image of the conductive strip.

## WJM0.10 used as functional ink

There is a growing research effort that focuses on flexible and printed electronics.[144] However, the real breakthrough is still to come, due to a number of technological challenges that need to be overcome. In particular, commercial printed electronics should be optically, electrically, and mechanically robust, with materials and components fulfilling basic performance criteria, such as low resistivity or transparency, under mechanical deformation.[8] Amongst the printing technologies, ink-jet printing is a promising technique for the direct deposition of nanomaterial-based inks.[145,146,147,148] We study the properties of the as-produced WJM0.10 used as ink for ink-jet printing.

In an ink-jet process, it is mandatory to obtain a stable jetting from the print-head nozzles. The stability of the jetting drop is dictated by various rheological properties such as density ($\rho$), surface tension ($\gamma$), and viscosity ($v$).[149] These properties, along with the nozzle size, need to be carefully tuned for the formation and ejection of droplets from the nozzle. In this context, the $Z$ number is commonly used as a FoM to control the ink quality in terms of regular drop formation, jetting accuracy, and attainable jetting frequency.[88,150] The $Z$ number is defined as the inverse of the Ohnesorge number $Oh=v(\gamma\rho\beta)^{-1/2}$, where $\beta$ is the printing cartridge nozzle diameter. If $Z$ is in the range $4<Z<14$ a good printing performance is expected to be guaranteed.[84] Specifically, Z values lower than 4 results in long-tailed droplet formation and Z values above 14 give rise to satellite drop formation.[151] Despite of this, several reports indicate that NMP-based inks can be used to print even at $Z$ values ~ 24.[152-153]

The viscosity of WJM0.10 (see Methods section for the detailed measurement procedure) is reported in Figure 9a. Considering that WJM0.10 has $v$ = 1.5 mPa s (from 1 to 100 s$^{-1}$), $\gamma$ = 41 mN m$^{-1}$ [153], $\rho$ = 1.3 g cm$^{-3}$, and the printing nozzle has a diameter $\beta$ = 21 µm, one calculates $Z$ = 20.9. This means that WJM0.10 is above the range for ink-jet printable inks.[152-153] Despite of this, neither satellite drops nor drop-tail are produced during the jetting of the ink from the cartridge. In fact, Figure 9b shows perfect WJM0.10 drops being ejected from the nozzle. The printability of the graphene based WJM-ink with $Z$ = 20.9 is demonstrated and is in agreement with Refs. 152-153. Patterns of 1×1 cm$^2$ on Si/SiO$_2$ (Figure 9c) have been printed. An image of the interconnected graphene flakes is shown in the SEM image, Figure 9d, demonstrating that the printing forms a continuous film. The sheet resistance R$_S$ of the printed electrode is 330 Ω□$^{-1}$ (with a thickness of 23 µ the conductivity is ~1.3 S cm$^{-1}$). This figure favourably compares with other results reported in the literature,[88,89,150,154] demonstrating that graphene obtained with WJM can be used as an ink-jet-able conductive ink.

## Conclusions

We have demonstrated the wet-jet milling as a method to produce large quantities of few-layer graphene dispersions, achieving concentration up to 10 gL$^{-1}$ with an exfoliation yield, *i.e.,* ratio between the weight of the processed material and the weight of the starting graphite flakes, of 100%. Our lab-scale set-up enables a production capability of up to 2.35 L h$^{-1}$. The average time required to produce one gram of exfoliated graphite is 2.55 min (23.5 g h$^{-1}$), which favourably outperforms other liquid-phase exfoliation processes such as ultrasonication, high-shear exfoliation, or microfluidization. The exfoliated flakes have a lateral size of ~460 nm and a thickness lower than 4 nm. Further purification, by ultracentrifugation of the as-produced WJM0.10 sample, promotes the enrichment of single-layer graphene. In fact, the percentage of single-layer graphene passes from ~10% in the as-prepared WJM0.10 sample to ~57% in the purified one. Additionally, we have shown the feasibility of wet-jet milling for the exfoliation of inorganic layered crystals, *i.e.*, hexagonal boron nitride, molybdenum disulphide, and tungsten disulphide, obtaining flakes with lateral sizes of 380, 500, and 340 nm, respectively.

The as-produced graphene flakes can be used without further purification for added-value applications. In particular, we have demonstrated the as-produced WJM0.10 as active material for anodes in lithium ion batteries, reaching 420 mAh g$^{-1}$; as filler in Polyamide–12 composites, getting an improvement of 34% of the flexural modulus; as ink-jet printable conductive ink, obtaining state-of-the-art electrical conductivity of ~1.3 S cm$^{-1}$.

## Acknowledgements

This work has been supported by the European Union's Horizon 2020 research and innovation program under grant agreement No. 696656—GrapheneCore1.

## Appendix

### Raman peaks assignment of graphene-graphite flakes

The Raman spectrum of graphene is composed by several characteristic peaks. The G peak, positioned at ∼1585 cm$^{-1}$, which corresponds to the E$_{2g}$ phonon at the Brillouin zone centre.[119,155,] The D peak, which is due to the breathing modes of the sp$^2$ hybridized carbon rings requiring a breaking on the carbon-ring symmetry for its activation by double resonance. Double resonance also happens as an intra-valley process, i.e., connecting two points belonging to the same cone around K or K',[119] resulting in the rise of the D' peak. The 2D peak (a second order resonance of the D band) centred at ∼2680 cm$^{-1}$ for an excitation wavelength of 514.5nm in case of a single layer graphene.[119] For few and multi-layer graphene the 2D peak is a superposition of multiple components, the main being the 2D$_1$ and 2D$_2$ components.[120] The 2D peak is always present, since no defects are required for the activation of two phonons with the same momentum, one being backscatter from the other.[119] In graphite the intensity of the 2D$_2$ band is roughly twice the 2D$_1$ band,[119] while for mechanically exfoliated single layer graphene (SLG) the 2D band is a single and sharp peak, which is roughly 4 times more intense than the G peak.[120]

## Exfoliation process reporting of size and thickness

As discussed in Ref. 119 the exfoliation of 2D crystals is considered as fragmentation process. This means the size distribution of the flakes follows a log-normal distribution. Following this model, all the lateral size and thickness reported in the main text corresponds to the distribution mode, which is the most frequent value in a data set (*i.e.*, the distribution peak).

The lateral sizes and thicknesses standard deviation are Log-normal standard deviations. Figure S1, shows the difference between the mode and the mean for a log-normal distribution.

## Surface tension-surface energy relationship

In order to exfoliate and stabilize a 2D crystal in a solvent, as stated in the main text, the Gibbs free energy of the mixture solvent/layered material should be minimized.[56,72,156,157] This condition can be fulfilled if the solvent surface tension is equivalent to the surface energy of the material by using the equation[64,72]

$$\gamma = E_{Surface}^{Solvent} - TS_{Surface}^{Solvent} \qquad \text{Eq.S1}$$

In which $E$ is the solvent surface energy, $T$ is the absolute temperature and $S$ is the solvent surface entropy which generally takes a value of $10^{-3}$ J m$^{-2}$ K$^{-1}$.[64,158]

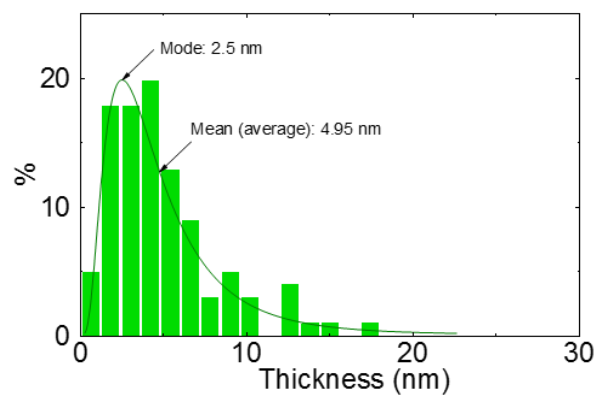

**Figure S1.** Log-normal distribution of exfoliated WS$_2$ flakes.

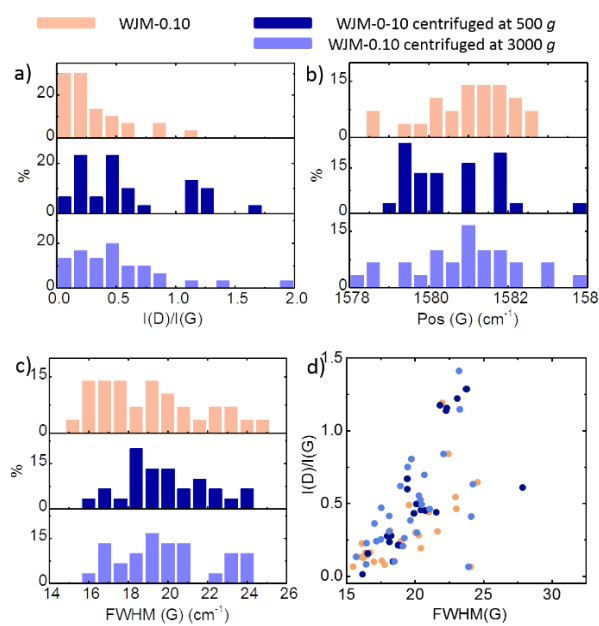

**Figure S2.** Statistical Raman analysis of the purified samples (dark and light blue or centrifuged at 500 and 3000 *g* respectively) compared with the as-produced WJM0.10 sample. a) I(D)/I(G), (b) Position of the G peak [Pos(G)]. c) full width at half maximum of G

[FWHM(G)] and d) Pos(2D).

## Raman analysis of purified samples

The statistics of the Raman spectra acquired on samples prepared at different centrifugation speeds is analysed in order to understand the morphological evolution of the graphite/graphene flakes during the purifications. The I(D)/I(G) statistical distribution broads when the centrifugal speed increases (Figure S2a), ranges from 0.03 to 1.2, for WJM0.10 samples, increasing to 2 for the WJM0.10-3000g. The G peak position and FWHM (G) are unchanged during ultracentrifugation with respect to the ones of the as-prepared sample. This indicates that no changes in doping and induced defects are present after the ultracentrifugation process (Fig S2 b-d).

## Exfoliation and characterization of other layered materials

### Optical absorption spectroscopy of MoS$_2$, WS$_2$ and h-BN.

The spectrum of MoS$_2$ is characterized by two excitonic bands, which are originating from the inter-band excitonic transition at the K point of the Brillouin zone and are located at 671 nm and 613 nm.[159,160,161] In the case of WS$_2$, the spectrum shows two main peaks around 530 nm and 639 nm corresponding to the excitonic absorption bands arising from the gap transition at the K point of the Brillouin zone.[162] The h-BN spectrum shows no bands in the wavelength range between 400 nm and 800 nm, but it is usually characterized by a band located at 236 nm corresponding to an optical band gap of ~5.26 eV, as it is shown in the reported literature on exfoliated h-BN nanosheets.[163] However, in this range there is the cut-off absorption of NMP (285 nm), thus making impossible to observe the transition peak.

The concentration of the dispersions obtained through UV-visible spectroscopy are reported in Table S1. All the extinction coefficients (α) and the reference wavelength (λ) used to calculate the concentrations are taken from the value reported in Ref. 72. The optical extinction coefficient is determined by using the Beer-Lambert law ($E = \alpha C l$, in which $E$ is the optical extinction at 600 nm, C the concentration of the exfoliated 2D crystal and $l$ is the path length, 0.01 m).

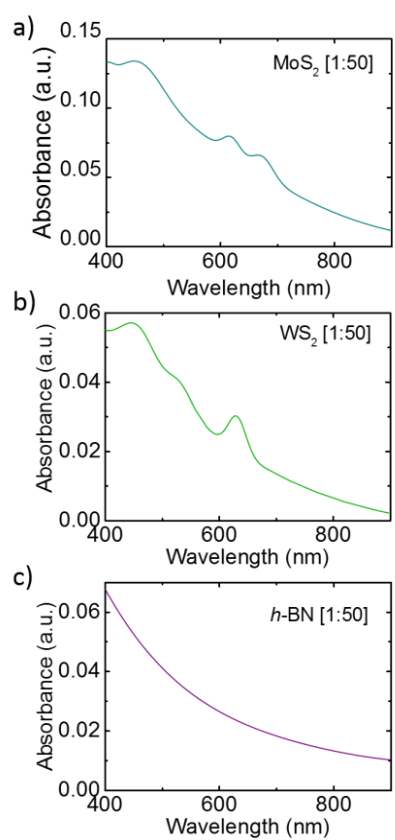

**Figure S3**. Optical absorption spectra of $MoS_2$, $WS_2$ and *h*-BN exfoliated by wet jet mill.

## Raman spectroscopy and statistics of $MoS_2$, $WS_2$ and *h*-BN.

The Raman spectra of bulk $MoS_2$ exhibits two main peaks, the first one the $A_{1g}$, located around 404.7 cm$^{-1}$ and $E_{2g}$ situated at 379 cm$^{-1}$.[123] The $A_{1g}$ is due to the in-plane vibration of the molybdenum and sulphur atoms, and the $E_{2g}$ corresponds to the out-of- plane vibration.[123] The shifting of the two peaks with respect to the bulk ones indicates that the number of layer has decreased. This behaviour has been discussed previously in literature.[122,164,165] Moreover, the peaks in the exfoliated material show a broadening with respect to the bulk peaks, which also indicates a decrease in the flakes thickness, as observed in the study of Ramakrishna Matte et al.[166]

| Material | α [Lm/g] | λ [nm] | C [g/L] |
|---|---|---|---|
| $MoS_2$ | 3400 | 672 | 0.05 |
| $WS_2$ | 2756 | 629 | 0.7 |
| *h*-BN | 2367 | 300 | 0.15 |

**Table S1.** Concentration and extinction coefficients.

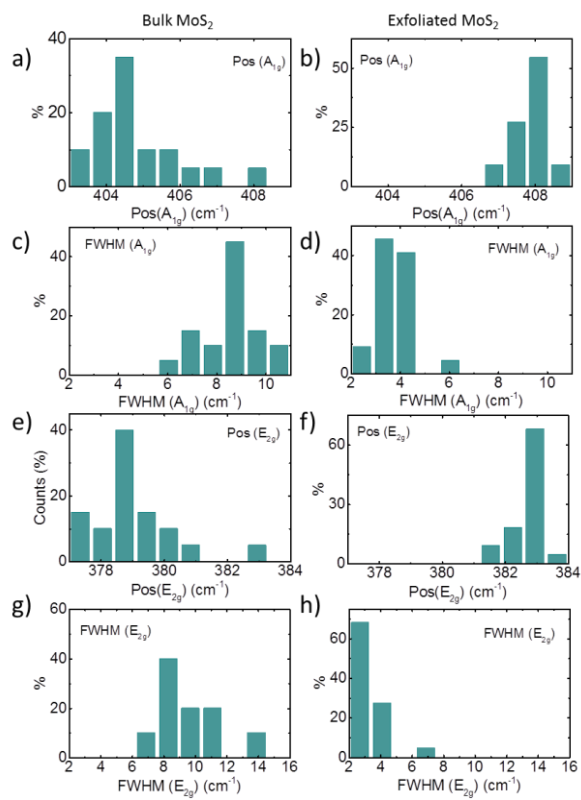

**Figure S4**. Raman statistical analysis of bulk and exfoliated $MoS_2$.

The bulk $WS_2$ Raman spectrum is characterized by two main peaks, $A_{1g}$ and $E_{2g}$, located around 418.9 cm$^{-1}$ and 353.6 cm$^{-1}$, respectively. The peaks corresponds to the out-of-plane mode related to the sulphur atoms and to the in-plane vibrational mode.[125,127] The difference between the bulk and the exfoliated material is given by a red shift of the peak $A_{1g}$, as, due to the fact that this peak is related to the interlayer interaction,[127] it is sensible to changes in thickness.[167]

In the previous studies about $WS_2$ nanosheets synthesis the $E_{2g}$ peak shows a blue shift in case of decreasing in number of layers.[127,168,169] In our case, there is an increase in FWHM($E_{2g}$), we can assume that the exfoliation has actually occurred, as the peak broadening is due to phonon confinement within the single layer.[167]

The bulk hexagonal boron nitride exhibits a characteristic peak situated at 1366 cm$^{-1}$ that is due to in-plane atomic displacements[128, 129,170] similar to the breathing mode associated to the G peak of graphene. The exfoliation of the *h*-BN from the bulk material to few layer or bilayer flakes is associated to a red shift of the peak E$_{2g}$.[129,171,163]

Thanks to the statistical analysis (performed on 106 flakes) it is possible to follow the general trend of the E$_{2g}$ peak position in the exfoliated and bulk *h*-BN. The maximum distribution of the Pos(E$_{2g}$) is down shifted 1 cm$^{-1}$, from 1366.6 to 1365.5 cm$^{-1}$, from the bulk to the exfoliated samples, respectively, demonstrating the thinning of the *h*-BN flakes.[163,129,171]

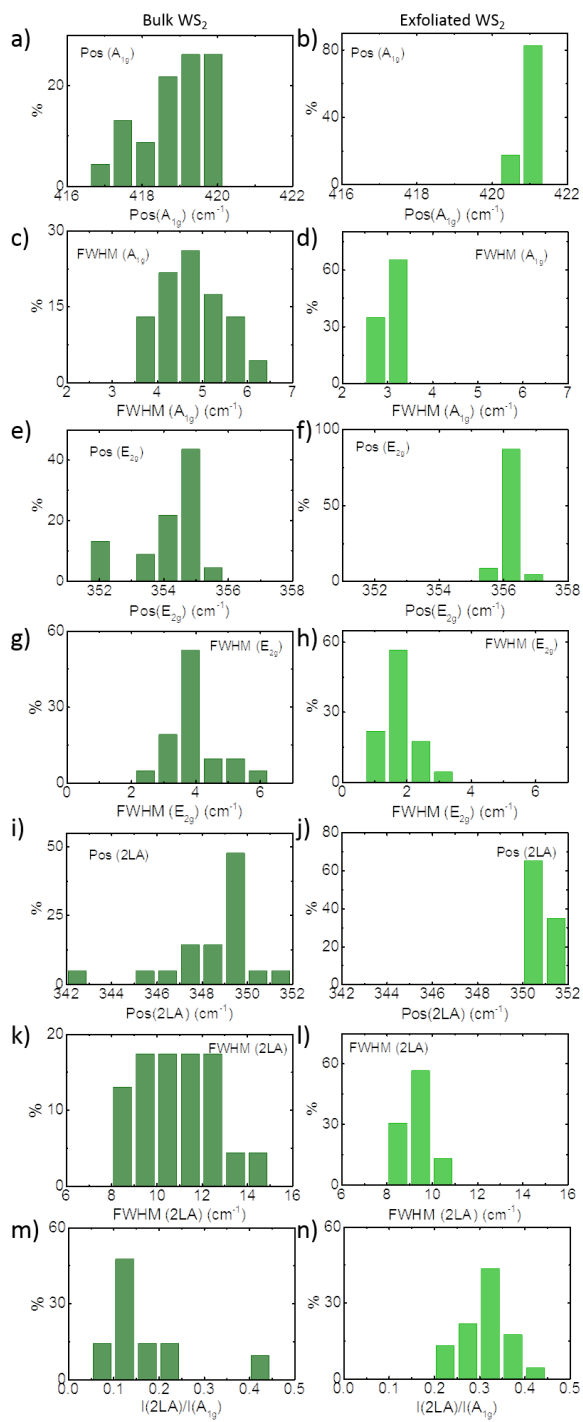

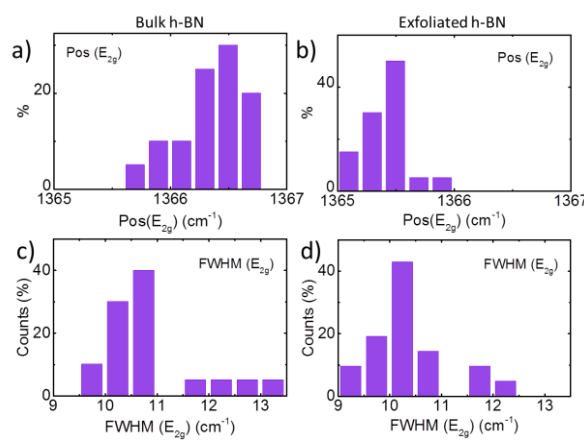

**Figure S6.** Raman statistical analysis of bulk and exfoliated *h*-BN.

**Figure S5.** Raman statistical analysis of bulk and exfoliated WS$_2$.